\documentclass[reprint,amsmath,amssymb,braket, aps]{revtex4-1}

\usepackage{graphicx,color}% Include figure files
\usepackage{dcolumn}% Align table columns on decimal point
\usepackage{bm}% bold math
\usepackage[colorlinks=true,linkcolor=blue]{hyperref}

\begin{document}

%\preprint{APS/123-QED}

\title{Plasmon damping in electronically open systems}% Force line breaks with \\

\author{Kirill Kapralov}
\affiliation{%
 Laboratory of 2d Materials for Optoelectonics, Moscow Institute of Physics and Technology, Dolgoprudny 141700, Russia}
\author{Dmitry Svintsov}%
 \email{svintcov.da@mipt.ru}
\affiliation{%
 Laboratory of 2d Materials for Optoelectonics, Moscow Institute of Physics and Technology, Dolgoprudny 141700, Russia}%


%\affiliation{
%\affiliation{
%\affiliation{%


\date{\today}% It is always \today, today,

\begin{abstract}
Rapid progress in electrically-controlled plasmonics in solids poses a question about effects of electronic reservoirs on the properties of plasmons. We find that plasmons in electronically open systems [i.e. in (semi)conductors connected to leads] are prone to an additional damping due to charge carrier penetration into contacts and subsequent thermalization. We develop a microscopic theory of such lead-induced damping based on kinetic equation with self-consistent electric field, supplemented by microscopic carrier transport at the interfaces. The lifetime of plasmon in electronically open {\it ballistic} system appears to be finite, order of conductor length divided by carrier Fermi (thermal) velocity. The reflection loss of plasmon incident on the contact of semi-conductor and perfectly conducting metal also appears to be finite, order of Fermi velocity divided by wave phase velocity. Recent experiments on plasmon-assisted photodetection are discussed in light of the proposed lead-induced damping phenomenon.

\end{abstract}

\pacs{Valid PACS appear here}% PACS, the Physics and Astronomy
\maketitle

Plasmons represent collective oscillations of charge carriers and electromagnetic field. Both instances can freely propagate in space. Free propagation of electromagnetic waves leads to the radiative decay of plasmons which has been studied extensively~\cite{Ritchie_rad_decay,Kokkinakis_rad_decay,Popov_rad_decay,Kukushkin_retardation_effects}. Charge carriers may also leak away from plasmonic system if it is coupled to electronic reservoirs (contacts). This process would also lead to plasmon damping. Unlike radiative damping, the 'contact damping' gained very little attention. The reason is most plasmonic systems studied so far were electronically closed (nanoparticles~\cite{Ritchie_rad_decay,Kokkinakis_rad_decay,Popov_rad_decay}) or extended~\cite{Kukushkin_retardation_effects,Koppens_low_loss,basov2018fundamental} such that effects of leads could be neglected. With the rapid progress in electrical control of plasmons~\cite{Koppens_Phase_shift,Stockman_spaser} and electrical readout of plasmon-enhanced photocurrent~\cite{Koppens_thermoelectric,bandurin2018resonant}, this question becomes urgent.

There are several experimental evidence for important role of contacts on plasmon damping that yet did not receive due attention. First, interference of launched and reflected plasmons is readily observed at edges of 2d and 1d semiconductors~\cite{Koppens_low_loss,Luttinger_plasmon_NT,Nonlinear_LL_plasmons}, but scarcely observed at contacts of semiconductors and metals~\cite{Koppens_Phase_shift}. Second, photocurrent spectroscopy of plasmon resonance in transistor structures with close leads provides generally larger linewidths~\cite{Knap_Resonant_detection,Luttinger_plasmon_NT_FET} compared to electromagnetic transmission measurements in grating-gated semiconductors with distant leads~\cite{Grating_gate_GaN_absorption}. In recent measurements of plasmon-enhanced photovoltage in graphene bilayer transistor~\cite{bandurin2018resonant}, the visibility of plasmon resonance was enhanced by $p-n$ junction barrier at the metal-graphene interface. These factors tell us that behaviour of plasmons at the contact of semiconductor and metal lead is not simply refection at impedance discontinuity.

\begin{figure}[ht!]
\center{\includegraphics[width=0.9\linewidth]{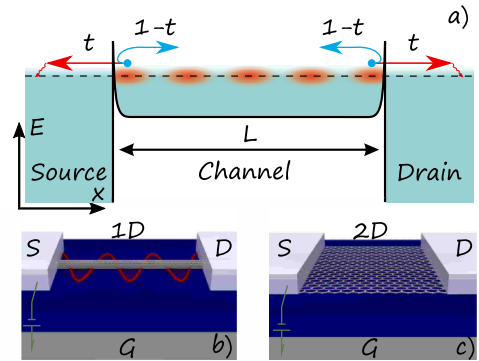}}
\caption{ 
(a) Electron occupation in a semiconductor channel coupled to metal leads vs energy and coordinate in the presence of plasma wave. Blue filled regions correspond to equilibrium electrons, red -- to non-equilibrium ones, excited by plasmon electric field. A non-equilibrium electron incident on a contact can either transmit and thermalize (with probability $t$) or be reflected back from a contact (with probability $1-t$) (b) Open plasmonic resonators with 1d and 2d channels (nanotube and graphene are shown as example) coupled to source and drain contacts}
\label{ris:Structure}
\end{figure}

The theory of plasmon decay in electronically open systems is still lacking. Its first theoretical evidence appeared in numerical Monte-Carlo simulations of plasmons in confined 2d systems with fixed electron distributions at contacts~\cite{Satou_contact_damping}. Further evidence appeared in simulations of current-driven plasmon instability~\cite{Mendl_instability}, though hardly distinguishable from bulk damping. Another approach to problem lies in finding the dynamic conductance of lead-coupled conductors and analysing its peaks as a function of frequency~\cite{Buttiker_Dynamic_Conductance}. Such calculation could be performed only for 1d systems with restrictive assumption of fully screened Coulomb interaction~\footnote{The boundary conditions on contacts cannot be posed in consistent way under assumption of fully screened Coulomb interaction. The local relation between potential $\varphi$ and electron density $n$, $C\varphi = e n$ leads to fixed electron density at the grounded contacts. This differs generally from microscopic boundary conditions derived from kinetic equation.}. Remarkably, the latter approach hints that plasmon damping in open systems is tightly linked to finite conductance of ballistic systems (given by Landauer~\cite{Landauer_1989} and Sharvin~\cite{sharvin} formulas in 1d and 2d, respectively).

In this Letter, we present an analytical theory of plasmon decay in 1d and 2d (semi)conductors coupled to leads~\footnote{The studied damping channel is also relevant for bulk (3d) plasmons, but its theoretical description if more complicated. The reason lies in degeneracy of 3d plasmon spectrum when spatial dispersion of conductivity is neglected.}. Schematic of this process is shown in Fig.~\ref{ris:Structure}: a non-equilibrium electron participating in plasma oscillation penetrates into a contact and is thermalized therein. We find that such damping appears due to non-locality of current-field response, i.e. it vanishes if Fermi velocity $v_{F}$ of carriers tends to zero. In the low-temperature limit, the damping $\gamma$ is order of $t v_F/L$, where $L$ is the distance between leads, and $t$ is the electron transmission probability at metal/semiconductor interface. An extra contribution to damping oscillatory in frequency with period $v_F/L$ is found. It appears due to synchronisation (anti-synchronisation) between periods of carrier transit and plasma oscillation. We also find that plasmons incident on semiconductor/metal contact undergo finite reflection loss due to the above mechanism. It occurs even for ballistic semiconductors and perfectly conducting metals.
%The amplitude of 'transit-time' oscillations in plasmon damping is most pronounced in lower dimensions. Particularly, in contact-induced plasmon damping can completely disappear in 1d systems at specific frequencies.

{ \it \label{sec:level1} Non-local conductivity in electronically-open system.} The main building block for evaluation of plasmon losses in open systems is the conductivity kernel $\sigma(x,x')$ linking the current density $j(x)$ and electric field $E(x)$
\begin{equation}
\label{eq-current-field}
j(x) = \int_0^L \sigma(x, x') E(x') dx'.    
\end{equation}
The non-locality of current-field relation (\ref{eq-current-field}) will play a central role in the effect studied. We are to find $\sigma(x,x')$ in $d$-dimensional semiconductor channel with metal leads located at $x=0$ and $x=L$ (Fig.~\ref{ris:Structure}). The metals are assumed perfectly conducting, and strong electron scattering maintains equilibrium Fermi distributions therein. Electron distribution in the channel obeys the classical kinetic equation; in the presence of plasmon electric field $ E(x) e^{-i\omega t}$ the distribution function $f(x,{\bf p}) = f_0({\bf p}) + \delta f (x,{\bf p})e^{-i \omega t}$ obeys
\begin{equation}
\label{eq:kinetic}
     -i \omega \delta f + v_x \frac{\partial \delta f}{\partial x} - e  E(x) \frac{\partial f_0}{\partial p_x} = 0,
\end{equation}
where $v_x$ is the $x$-component of electron velocity. We shall focus on ballistic systems with long momentum relaxation time $\tau_p\omega \gg 1$ to clearly distinguish the bulk and contact damping.

A non-equilibrium electron incident on a contact can either penetrate and thermalize therein with probability $t$, or undergo specular reflection from metal-semiconductor junction with probability $r = 1-t$. These considerations relate the distributions of left- and right-moving electrons at the contacts:
\begin{equation}
\begin{split}
 \delta f(0, p_x) = r \ \delta f(0, - p_x), \\
   \delta f(L, -p_x) = r \ \delta f(L, p_x).
 \end{split}
\end{equation}

Solving the kinetic equation (\ref{eq:kinetic}) at given field $\delta E(x)$, we have obtained the nonlocal conductivity kernel of the form
\begin{equation}
\label{eq-Conductivity-kernel}
\sigma(x, x') = \sigma_{D} d \frac{i \omega}{2 v_F}  \left\langle F\left(\frac{x}{L}, \frac{x'}{L}, \Omega_{\theta}\right) \cos\theta \right\rangle_{\cos\theta >0},
\end{equation}
\begin{multline}
F(\xi, \eta, \Omega_{\theta}) = e^{-i \frac{|\xi -\eta|}{\Omega_{\theta}}} -\\
2r \frac{\cos[\frac{\xi+\eta-1}{\Omega_{\theta}}] - r \cos[\frac{\xi-\eta}{\Omega_{\theta}}]e^{-i/\Omega_{\theta}}}
{e^{i/\Omega_{\theta} } -  r^2 e^{-i/\Omega_{\theta}}}    
\end{multline}
where $\sigma_{D} = i n_e e^2 / \omega m$ is the local Drude conductivity, $n_e$ is the electron density, $\Omega_\theta = v_F \cos\theta/\omega L$ is the normalized transit frequency of charge carrier moving at angle $\theta$, and the angular averaging $\langle...\rangle_{\cos\theta >0}$ is performed over right-moving carriers. It is possible to show that the local Drude form of conductivity $\sigma(x,x') \approx \sigma_{D} \delta(x-x')$ is restored in the limit of small transit frequency $\Omega_{t} = v_F/\omega L \rightarrow 0$, i.e. at high frequencies and low Fermi velocities.

{\it Damping rate in electronically open resonator}. To quantify the damping rate of plasmons $\gamma_{\rm cont} = -{\rm Im}\omega$ induced by coupling to leads, we divide the energy loss rate in open plasmonic resonator $dW/dt$ by the stored energy $W$:
\begin{equation}
\label{Energy-loss}
\frac{\gamma_{\rm cont}}{\omega_n} = 
\frac{1}{2} \frac{\int_0^L E^*(x')\sigma'(x,x') E(x)\,dxdx'}{ \int_0^L{\sigma''_D|E(x)|^2dx}}
\end{equation}
where the prime and double prime denote real and imaginary parts of conductivity.

The possibility to express contact-induced damping in the from of bulk Joule loss (\ref{Energy-loss}) is non-trivial, yet it can be formally derived from perturbation theory for coupled dynamic and electrostatic equations~\cite{Perturbation_Theory}. Actually, all information about electron reflection/absorption at contacts is now encoded in conductivity kernel $\sigma(x,x')$. Energy loss associated with direct leakage of carrier kinetic energy through the contacts should also contribute to $\gamma_{\rm cont}$. However, this flux is order of $(\omega/q) m v_F^2/2$; we will see that actual damping rate given by (\ref{Energy-loss}) appears already to the lower order in Fermi velocity.

%Upon derivation of Eq. (\ref{Energy-loss}) we have used (1) equality of average potential energy of charge interactions and average kinetic energy of carriers (2) negligible energy flux through the contacts, to the leading order in non-locality $\Omega_t$.

\begin{figure}[ht!]
\center{\includegraphics[width=1\linewidth]{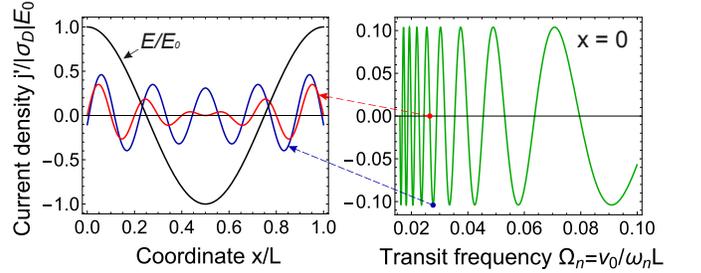}}
\caption{ 
  Spatial distribution of current ${\rm Re}j(x)$ in 1D conductor coupled to perfectly transmitting leads ($r=0$), induced by  a trial plasmon field $E_0\cos(2\pi n x/L)$. To visualize the localization of current, we have added a small imaginary part of frequency mimicking the internal scattering. The currents are periodic in space with wave vector $\omega/v_F$, and their amplitude (panel b) oscillates with transit-time frequency} 
\label{ris:Current}
\end{figure}

The dissipation of plasmons is tightly linked to the hermiticity of conductivity kernel $\sigma(x,x')$, as it is seen from Eq.~(\ref{Energy-loss}). First of all, the kernel has vanishing real part for perfectly reflecting contacts $r \rightarrow 1$. This implies absence of dissipation once charge carriers cannot escape the channel. Second, the matrix elements of $\sigma(x,x')$ evaluated on plane-waves $e^{ik_x x}$ in infinite volume $L\rightarrow \infty$ are zero if the phase velocity $\omega/k_x$ exceeds the Fermi velocity. This implies the absence of Landau damping for sufficiently fast waves. Once the transmittance of contacts and channel length become finite, the conductivity acquires dissipative part, ${\rm Re}\sigma(x,x')>0$, and the contact-induced damping appears.

%To obtain the plasmons' eigen frequencies in open resonator, we supplement the non-local transport relation with Poisson's equation. The latter couples the self-consistent electric potential $\varphi(x)$ and induced charge density $\rho(x) = (i \omega)^{-1} \partial j/\partial x$:
%\begin{equation}
%\label{eq-Poisson} 
%    \varphi(x) = - \frac{4 \pi}{i \omega} \int\limits_0^L G(x,x') \frac{\partial j(x')}{\partial x'} dx'.
%\end{equation}
%Above, $G(x,x')$ is the Green's function of Poisson's equation governed by the geometry of metal leads and gates.

%A key observation enabling us to find contact contribution to plasmon damping is the reality of eigen-frequencies in the local limit, $v_F/\omega L \ll 1$. Physically, the electrons are can leak to contacts at the length scale $l_T = v_F/\omega$, which is the electron transit length during plasma oscillation period. When $l_T$ is small compared to resonator length $L$, most electrons behave as if they were bulk. Therefore, the solutions of coupled transport and Poisson's equation in the local limit (frequencies $\omega^{(n)}_{loc}$ and potentials $\varphi^{(n)}_{loc}$) can be used as a good starting point to construct perturbation theory with respect to contact effects. Such local solutions are readily obtained with available electromagnetic simulators.

To provide a link between contact-induced damping and Landau damping, we inspect the spatial profile of current (\ref{eq-current-field}) induced by a 'trial' long-wavelength field $E \sim \cos (\pi n x/L)$. We observe a short-period structure in current localized in the vicinity of both contacts, as shown in Fig.~{\ref{ris:Current}}A. The amplitude of these 'lead-induced' currents oscillates with electron transit frequency (Fig.~2B), while the phase shift with respect to electric field tends to $\pi/2$ for perfectly reflecting contacts ($t=0$).
%\addDS{With $r = 0$ it has the form: }
%\begin{multline}
%\label{Current}
%j(x) =  j_0(x) - \\ \frac{1}{2} \left\langle \left( j_0(0) e^{-\frac{i \omega}{v_x}x} + j_0(L) e^{-\frac{i \omega}{v_x}(L - x)} \right) \cos{\theta}  \right\rangle_{\theta > 0} 
%\end{multline}

%\addDS{ In the 1D case it is possible to show, that the edge current density goes to zero with $\Omega_t = (\pi 2 k)^{-1}$ for even modes and $\Omega_t = (\pi (2k-1))^{-1}$ for odd modes.  }
The appearance of such structures is a consequence of non-local conductivity only, and is not related to electric field enhancement at sharp metal contacts. It is also possible to show that the phase of short-period current coincides with that of external field, which leads to a {\it spatially localized} Landau damping.

%It is feasible if the vertical extent of the leads much exceeds the plasmon wavelength. The potential distributions and frequencies in this case are simply
%\begin{equation}
%    \varphi^{(n)}_{loc} \propto \sin(q_n x),\qquad \omega^{(n)}_{loc} = \omega_\infty(q_n)
%\end{equation}
%where $\omega_\infty(q)$ is the frequency-wave vector plasmon dispersion in extended system, while the quantization rule is $q_n = \pi n/L$. To the leading orders of perturbation theory with respect to normalized transit frequency $\Omega_t$, the contact correction to eigen-frequency can be presented as   

We now quantify the contact contributions to plasmon damping in several experimentally relevant structures, such as (a) 1d nanotube field-effect transistor (FET) (b) FET with gated two-dimensional channel. If the vertical extent of leads is well above plasmon wavelength, the quantized frequencies of plasmon resonances are approximately $\omega_n = \omega_\infty(q_n)$, $q_n = \pi n/L$, where  $\omega_\infty(q)$ is the plasmon dispersion in extended structure without contacts, and $q$ is the wave vector. The electric potential of plasmon in these structures, in the non-retarded approximation, is $\varphi^{(n)} \propto \sin(q_n x)$~\cite{Exact}. We plug the electric fields $E_n(x) = -\partial\varphi^{(n)}/\partial x$ into (\ref{Energy-loss}) to obtain the following estimates of damping $\gamma_{\rm cont}$:

 \begin{figure}[h]
\center{\includegraphics[width=1\linewidth]{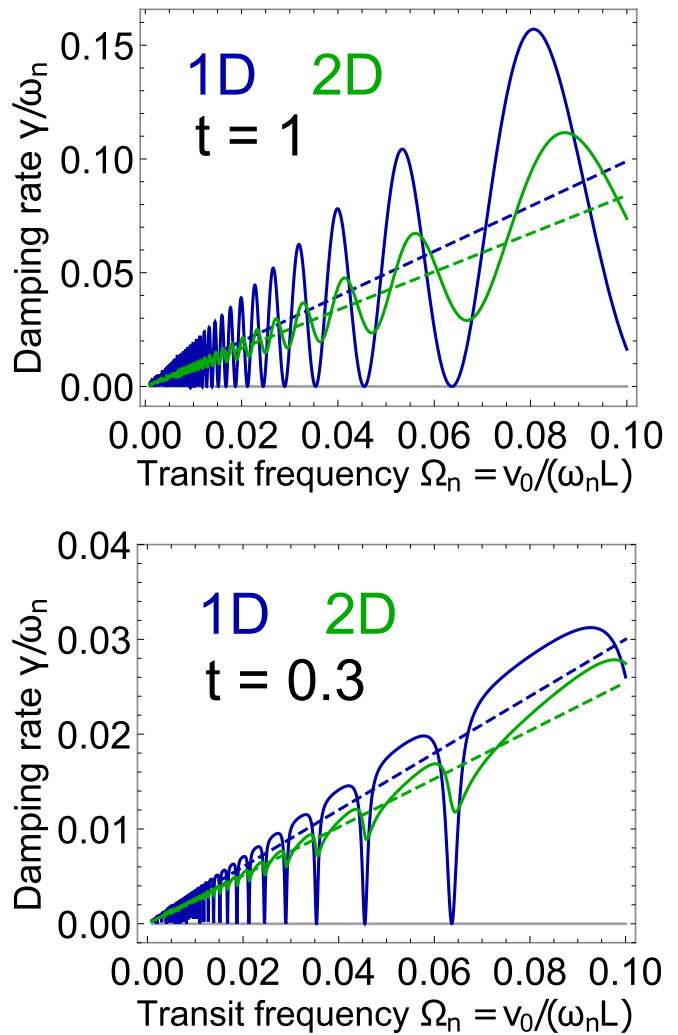}}
\caption{Plasmon damping rate $\gamma$ (normalized by eigen-frequency $\omega_n$, $n=1$) in ballistic semiconductor channel coupled to leads vs electron transit frequency. Blue and green lines correspond to 1d and 2d semiconductor channels, respectively; upper pannel corresponds to perfectly transmitting (ohmic) contact, lower panel -- to contact with transmittance $t=0.3$. Dashed lines show the non-oscillatory part of damping being linear in transit frequency.}
\label{ris:Dampig}
\end{figure}

\begin{gather}
\label{main}
     \frac{\gamma_{\rm cont}}{\omega_n} = - \alpha(d) \ t \Omega_t - \beta(d) \ t^2 \Omega_t^{(d+1)/2} \Phi_{\rm osc}(\Omega_t),  \\
    \Phi_{\rm osc}(\Omega_t)= \sum_{k=1}^{\infty} \frac{r^{k-1}}{k^{\frac{d-1}{2}}}   \cos\left[ \frac{k}{\Omega_t} + \frac{\pi}{4}(d-1) + \pi n k \right]
\end{gather}
The numerical prefactors $\alpha$ and $\beta$ depend on channel dimensionality and are given by:
\begin{equation*}
\begin{split}
     \alpha(1) = 1; \ \alpha(2) = \frac{8}{3 \pi}; \
    \beta(1) = 1; \ \beta(2) = \sqrt{\frac{8}{\pi}}.
\end{split}
\end{equation*}

The contribution to damping given by first term of (\ref{main}) is the mean lifetime of a free electron in the channel of length $L$ confined by barriers with transparency $t$. The coefficient $\alpha$ decreases in higher dimensions due to longer transit time of electron between source and drain, {\it averaged} over the Fermi surface. 

The second term of the equation (\ref{main}) appears only in structures with partially reflecting contacts, and describes possible resonances (anti-resonances) between electric field and bouncing electrons~\cite{Ballistic_impedance}. An enhancement of damping occurs if plasmon field acts in phase with most of charge carriers, while its reduction occurs when the field and carriers oscillate in counter-phase. 

The interaction between plasmons and transit-time electron oscillations is most illustrative in one-dimensional channels, where the damping rate is found fully analytically 
    \begin{equation}
       \left( \frac{\gamma_{\rm cont}}{\omega_n}\right)_{d=1} = -\Omega_t t \frac{1+r}{r} \dfrac{1 - (-1)^n\cos \Omega_t^{-1} }{[1-(-1)^n\cos \Omega_t^{-1}] + t^2/2r} 
    \end{equation}
One-dimensional nature of electrons in this case leads to a very strong interaction between field and transit-time resonances. The damping rate of plasmons goes to zero at $\Omega_t = (2 \pi k)^{-1}$ for even modes and $\Omega_t = (\pi (2k-1))^{-1}$ for odd modes, and so does the current at the terminals. When contact transmittance is low $t \ll 1$, the suppression of electron spill-out has a resonant character with width $\delta \omega \sim t \frac{\omega_{pl}}{q_k L}$.
%In higher dimesions we always have electrons, which moves at an angle to the x-axis and full leak suppression becomes impossible. This is the same frequency dependence as the admittance of transit-time diode. 

{\it Energy loss of a wave incident at contact}. The contact mechanism of damping would also manifest itself in plasmonic interference phenomena near the metal-semiconductor contacts. Analysis of such interference patterns is an established tool for determination of spectra and propagation length of plasmons~\cite{Luttinger_plasmon_NT,Koppens_low_loss}. 

We now realize that plasmon reflection from metal-semiconductor contact can never be perfect as some fraction of carriers would penetrate into metal and thermalize therein. Attenuated reflection from such contact is a consequence of non-local conductivity; no attenuation would have occurred if semiconductor was described by a local Drude conductivity.

The method for calculation plasmon reflection loss $A_{\rm pl} = 1 -R_{\rm pl}$ at semiconductor-metal contact (located at $x=0$) is similar to evaluation of damping in confined structure. Namely, it equals the Joule losses (expressed through non-local conductivity) divided by energy flux in an incoming wave $S_{\rm inc}$:
\begin{equation}
\label{scattering-absorption}
    A_{\rm pl} = \frac{1}{S_{\rm inc}}\int_{-\infty}^0{ E^*(x')\sigma'(x,x') E(x)\,dxdx'},
\end{equation}
The last necessary element for evaluation of losses is conductivity kernel in a semi-infinite semiconductor channel $\sigma(x,x')$. It is related to non-local conductivity of extended system $\sigma_\infty (x-x')$ via $\sigma(x,x') = \sigma_\infty (x-x')  - r \sigma_\infty (x + x')$~\cite{reuter1948theory}. It can be obtained from conductivity in finite-length channel (\ref{eq-Conductivity-kernel}) by taking the limit $L\rightarrow \infty$ and keeping in mind small wave damping. Finally, taking the zero-loss form of incident field $E = E_0 \cos(q x)$ and evaluating the integrals, we find the reflection loss in one and two dimensions:
\begin{equation}
    A_{pl} = \xi(d) t \frac{v_F}{\omega / q},
\end{equation}
where $\xi(1) = 1 $,  $\xi(2) = 16/3$. The result is remarkably simple: reflection loss of a plasmon incident on metal contact is the ratio of carrier Fermi and plasmon phase velocities, timed by the transmission coefficient for individual carrier $t$. Of course, leading term of plasmon damping in open resonator (\ref{main}) is consistent with the obtained reflection loss.

\begin{figure}[h]
\center{\includegraphics[width=0.9\linewidth]{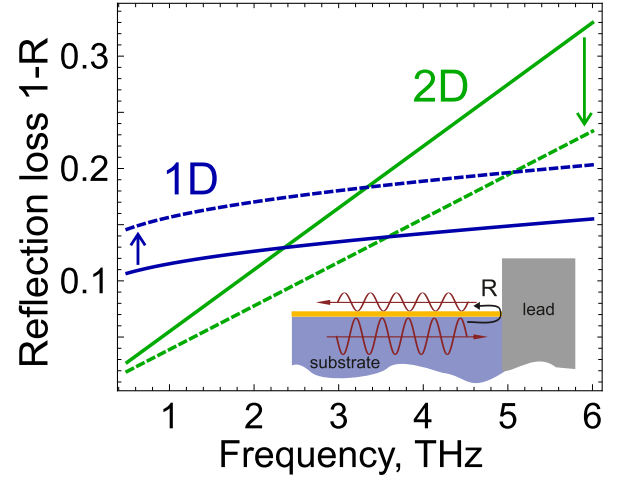}}
\caption{Plasmon reflection loss upon scattering at semiconductor/metal contact vs wave frequency. Blue and green lines correspond to 1d (semiconductor nanotube~\cite{Nonlinear_LL_plasmons}) and 2d (GaAs quantum well) plasmons; solid lines correspond to carrier densities $n = 5 \times 10^5$ cm$^{-1}$ in 1d and $5 \times 10^{11}$ cm$^{-2}$ in 2d; dashed lines -- to higher densities of $10^6$ cm$^{-1}$ in 1d and $10^{12}$ cm$^{-2}$ in 2d.
The radius of 1d conductor is assumed $a = 5$ nm, effective mass $m^* = 0.1 m_0$, background dielectric constant $\kappa = 4$. Inset shows a schematic of lossy plasmon reflection at the contact
}
\label{ris:ReflectionLoss}
\end{figure}

It is possible to express the reflection losses (\ref{scattering-absorption}) via wave vector or via frequency only. This enables one to estimate the relative smallness of such contact absorption, and results in following relations
\begin{equation}
    A_{\rm pl}^{(1D)} =  t \sqrt{\dfrac{\pi}{2 \alpha_c} } {\rm ln}^{-1/2}  \frac{v_F/a}{\omega}  ; \ \ A_{\rm pl}^{(2D)} =  \frac{16}{3} \frac{t \omega}{v_F/a_B} ,
\end{equation}
where $\alpha_c = e^2 / \kappa \hbar v_0$ is the Coulomb coupling constant, $a$ is the width of one-dimensional system (e.g. radius of the nanotube), and $a_B = \hbar^2 \kappa/me^2$ is the Bohr's radius. It is remarkable that reflection losses in 1d and 2d have the opposite dependencies on carrier density, as shown in Fig.~\ref{ris:ReflectionLoss}. In 2d, increase in carrier density leads to larger phase velocity, smaller non-locality, and smaller loss. In 1d, the increase in Fermi velocity with carrier density overtakes that of wave phase velocity, and reflection loss goes up with increasing density.

{\it Discussion and possible experimental manifestations.} Lead-induced damping can make a contribution to net plasmon damping in any electrically-controlled plasmon resonator (e.g., in experiments with gate tuning of plasmons). The most pronounced effect is expected in plasmon-enhanced photodetectors, where a semiconductor channel acts as a plasmonic resonator and photocurrent generator~\cite{Dyakonov_Shur,Luttinger_plasmon_NT_FET,Knap_Resonant_detection}. Plasmon lifetime in such detectors based on bilayer graphene was inferred in Ref.~\cite{bandurin2018resonant} from the width of gate-tunable photovoltage oscillations. The extracted lifetime $\gamma^{-1}\sim 0.3...1$ ps was well below the double transport relaxation time ($\sim 4$ ps) predicted by Boltzmann kinetic theory in extended 2DES. Moreover, the lifetime was decreasing at larger gate voltage (corresponding to higher Fermi velocities). The latter trend is in agreement with Eq.~(\ref{main}) for contact damping.

The magnitude lead-induced damping under conditions of Ref.~\cite{bandurin2018resonant} can be estimated using Eq.~(\ref{main}) as $16 v_F/3\pi L \approx 2.7 \times 10^{11}$ s$^{-1}$ (taking $v_F = 10^6$ m/s and $L = 5$ $\mu$m). It is roughly the same as 'bulk damping' $1/2\tau_p = 2.5 \times 10^{11}$ s$^{-1}$. Yet, the net damping rate is even above $\gamma_{\rm cont} +(2\tau_p)^{-1}$, which signalizes on extra plasmon decay mechanisms (radiative decay is a likely candidate). Remarkably, measurements of plasmon-assisted resonant detection in sub-micron III-V transistors ($L = 150$ nm)~\cite{Knap_Resonant_detection} also reported plasmon lifetime $\sim 200$ fs smaller than expected 800 fs from mobility measurements. The magnitude of measured lifetime is close to $v_F/L \sim 150$ fs expected for contact mechanism.

It looks that contact mechanism of plasmon damping sets a limit for downscaling of plasmonic nanosystems. Due to formal coincidence of boundary conditions at perfect contact and rough metal surface~\cite{reuter1948theory}, this damping would persist for localized plasmons in nanoparticles and surface plasmon-polaritons at metal/dielectric interface~\cite{Surface_scattering}. A straightforward way for reduction of such damping is to induce weakly transparent barriers for electrons near the contacts. One may therefore state that plasmonic devices should benefit from high {\it contact resistance} and {\it long channels}, which is contrary to requirements for conventional high-frequency transistors. Another possible way to reduce contact damping is to shift the carrier transport into hydrodynamic regime. In this regime, the path of a single carrier to the contact would be prolonged due to frequent electron-electron collisions~\cite{gurzhi1968contribution}, and so will the plasmon lifetime. However, the study of this regime requires a considerable modification of transport model.

To conclude, we have shown that coupling of a semiconductor system to metal leads induces extra plasmon damping. The damping appears due to electron penetration into leads and subsequent thermalization. The contribution of this mechanism to damping rate in semiconductor of length $L$ is roughly $v_F/L$; the contribution to plasmon reflection loss at semiconductor/metal interface is roughly the ratio of $v_F$ and wave phase velocity. 

The authors thank Aleksandr S. Petrov for rigorous derivation of Eq.~(\ref{Energy-loss}) and Denis Bandurin for stimulating discussions. The work was supported by the grant \# 16-19-10557 of the Russian Science Foundation.

\bibliography{Bibliography}

\end{document}